\documentclass[ reprint, aps, prl, two column, showkeys, groupedaddress]{revtex4-1}
\usepackage{graphicx,epsfig}
\usepackage{amssymb}
\usepackage{amsmath}
\usepackage{bm}
\usepackage{textcomp}
\usepackage{soul,xcolor}

\usepackage{color}

\begin{document}
\setcounter{page}{1}
\setstcolor{red}

\title[]{Surface segregation and the Al problem in GaAs quantum wells}
\author{Yoon Jang \surname{Chung}}
\author{K. W. \surname{Baldwin}}
\author{K. W. \surname{West}}
\author{M. \surname{Shayegan}}
\author{L. N. \surname{Pfeiffer}}
\affiliation{Department of Electrical Engineering, Princeton University, Princeton, NJ 08544, USA  }
\date{\today}

\begin{abstract}

Low-defect two-dimensional electron systems (2DESs) are essential for studies of fragile many-body interactions that only emerge in nearly-ideal systems. As a result, numerous efforts have been made to improve the quality of modulation-doped Al$_x$Ga$_{1-x}$As/GaAs quantum wells (QWs), with an emphasis on purifying the source material of the QW itself or achieving better vacuum in the deposition chamber.  However, this approach overlooks another crucial component that comprises such QWs, the Al$_x$Ga$_{1-x}$As barrier. Here we show that having a clean Al source and hence a clean barrier is instrumental to obtain a high-quality GaAs 2DES in a QW. We observe that the mobility of the 2DES in GaAs QWs declines as the thickness or Al content of the Al$_x$Ga$_{1-x}$As barrier beneath the QW is increased, which we attribute to the surface segregation of Oxygen atoms that originate from the Al source. This conjecture is supported by the improved mobility in the GaAs QWs as the Al cell is cleaned out by baking.
\end{abstract}
\maketitle

High-quality, single-crystal materials are ideal hosts for two-dimensional electron systems (2DESs) when studying electron-electron interaction driven phenomena. A typical example is the case of modulation-doped GaAs/Al$_x$Ga$_{1-x}$As heterostructures, where exotic states such as the fractional quantum Hall phase were first observed \cite{Tsui.PRL}. It has also been reported that Si \cite{Nelson.APL.1992,Kott.PRB.2014,Lu.PRB.1970}, AlAs \cite{Lay.PRB.1993,DePoortere.APL.2002,Shayegan.PSS.2006,Chung.PRM.2017}, GaN \cite{Manfra.JAP.2002}, ZnO \cite{Tsukazaki.Nat.2010}, Ge \cite{Shi.PRB.2015}, graphene \cite{Du.Nat.2009,Bolotin.Nat.2009,Dean,Feldman.PRL.2013} and InAs \cite{Meng} qualify as excellent candidates for investigation of electron-electron interaction phenomena. Each of these materials portray distinct electronic properties, and great endeavors are being made to improve their quality to fully understand the influence of such characteristics on many-body interactions. 

Perhaps because of the close matching of the lattice constants of the GaAs quantum well (QW) and the Al$_x$Ga$_{1-x}$As barrier material, as well as the simplicity of the GaAs conduction-band minimum resembling a free-electron model, historically GaAs 2DESs have been the focus of much effort to improve sample quality. As a result, modern molecular beam epitaxial (MBE) grown GaAs 2DESs were the first to show delicate many-body ground-states such as stripe/bubble phases \cite{stripebubble}, Wigner crystal \cite{WignerRev,Andrei,Jiang,Wigner}, and the $\nu=5/2$ fractional quantum Hall state \cite{fivehalf}, and they continue to be a leader in revealing new phenomena. Given these achievements, the motivation to push for even cleaner GaAs samples is strong, as there is potential for observing new and exciting emergent quantum phenomena. Indeed, various avenues are being pursued to achieve this goal \cite{ManfraReview}, with a recent emphasis on the purification of the Ga source \cite{MGa,GaProblem}. Although this is a crucial effort  considering that Ga is what actually comprises a GaAs QW where the 2DES resides, it may overlook another component that is inevitable in state-of-the-art modulation-doped structures: the Al in the Al$_x$Ga$_{1-x}$As barrier. 

It is well known that when intentional dopants such as Si, Sn, or Be are introduced to GaAs or Al$_x$Ga$_{1-x}$As layers during the MBE growth, surface segregation is a significant issue \cite{AlCho,Heiblum,SantosAPL,Lanzilotto,Lanzilotto2,Schubert}. This is also the case with unintentional impurities such as O due to its high reactivity with the Al in the Al$_x$Ga$_{1-x}$As as previous secondary ion mass spectrometry (SIMS) data have shown \cite{SIMS.JCG}. We can then expect that some of the O accumulated in the barrier during growth will migrate toward the GaAs in a QW structure. Indeed, this effect has been observed in the past from SIMS measurements and has been conjectured to be a crucial factor in determining the quality of an Al$_x$Ga$_{1-x}$As/GaAs interface \cite{MarcIllegems}. Here, we scrutinize this view by conducting a series of experiments to quantify the influence the surface segregation of impurities from the barrier has on the mobility of a high-quality GaAs QW ($\mu\simeq10^7$ cm$^{2}$V$^{-1}$s$^{-1}$).  From high-resolution SIMS results, we also show that the O in the barrier not only migrates toward the GaAs QW but is actually discharged into it near the Al$_x$Ga$_{1-x}$As/GaAs interface. Our results clearly demonstrate that having a clean Al source which prevents this disposal of O into the QW is crucial for achieving high-mobility GaAs samples, especially when having thick and/or high $x$ barriers in the structure. 

\begin{figure} [t]
  \begin{center}
    \psfig{file=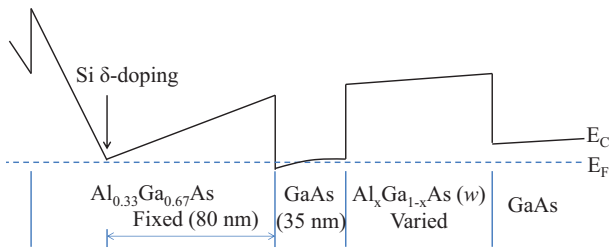, width=0.45\textwidth }
  \end{center}
  \caption{\label{fig1} Schematic diagram depicting the conduction-band edge energy ($E_{C}$) and the Fermi energy ($E_{F}$)for the sample structure used in this study. All samples are only doped in the top-side (Al$_{0.33}$Ga$_{0.67}$As) barrier with a spacer thickness of 80 nm to fix the 2DES density at $\simeq2.1\times10^{11}$ cm$^{-2}$. The well width is also fixed to be 35 nm except for the case where there is no underlying Al$_x$Ga$_{1-x}$As layer. The two variables in this study are the width of this underlying Al$_x$Ga$_{1-x}$As layer ($w$) and its alloy fraction ($x$).}
\end{figure}

For our study, we grew several modulation Si $\delta$-doped GaAs QWs by MBE (see Fig. 1). Only the top side of the structure was doped, where the Al$_x$Ga$_{1-x}$As barrier alloy fraction and spacer thickness were fixed at $x=0.33$ and $s=80$ nm for all samples. To evaluate the influence of surface segregation from the barrier on the quality of the GaAs QW, we varied the thickness ($w$) of the underlying barrier from $w=0$ (simple heterostructure) to $w=350$ nm, while fixing the barrier alloy fraction at $x=0.33$, and measured the transport mobility of the 2DES in the GaAs QWs at $T=0.3$ K. The width of the GaAs QW was fixed at 35 nm except for the heterostructure case ($w=0$). We repeated this process as we outgassed and purified our Al cell after a new batch was loaded into our MBE chamber \cite{fnote10}. We also investigated the impact of barrier alloy fraction by implementing the same structure described in Fig. 1 and varying $x$ from  0 (heterostructure) to 1 (pure AlAs barrier). In this case, we grew two series of samples, one with $w=10$ nm and the other with $w=200$ nm, to differentiate the effect of merely introducing a barrier and defining a QW from surface segregation. 

All electrical measurements were performed with low-frequency lock-in amplifiers on samples with van der Pauw geometry in a $^{3}$He cryostat. For these measurements, the samples were first cooled down to $\simeq 10$ K and then exposed to red LED illumination for $\simeq 5$ minutes. The LED was then turned off and the samples were cooled to 0.3 K. The cooling procedure was fixed for all samples. We evaluated the 2DES density in the QWs from the integer/fractional quantum Hall features in the magnetoresistance data. For SIMS measurements, we prepared samples without the doped top barrier to prevent any contributions from the top Al$_{0.33}$Ga$_{0.67}$As layer via the knock-on effect \cite{knockon}.  

\begin{figure}
  \begin{center}
    \psfig{file=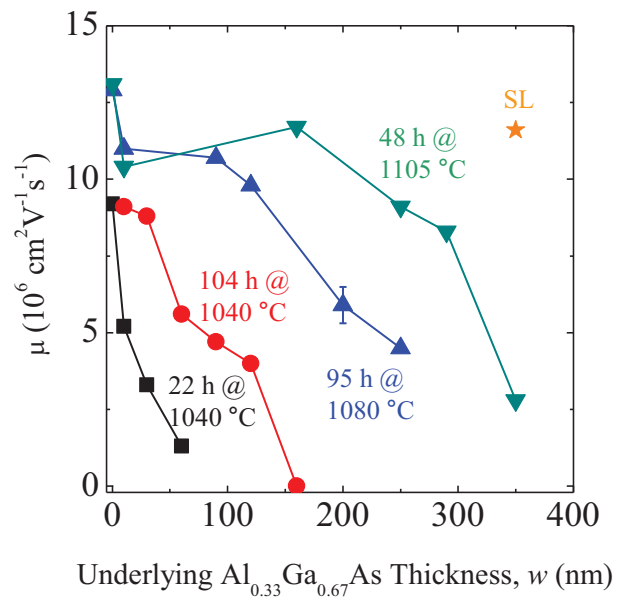, width=0.45\textwidth }
  \end{center}
  \caption{\label{fig2} Measured mobility values of the 2DES in GaAs QWs as a function of the underlying Al$_{0.33}$Ga$_{0.67}$As barrier thickness. The deviation of the measured values in different pieces of the same wafer was roughly $10\%$. A representative error bar is shown for the case of a 200-nm barrier in the blue profile. Each colored profile denotes data acquired after a bake-out of the Al cell, specified by text in the same color. All bake-outs were performed on the same Al cell in a cumulative fashion. The data point SL represents the case where a 350-nm-Al$_{0.33}$Ga$_{0.67}$As barrier was replaced with a 10-nm-Al$_{0.33}$Ga$_{0.67}$As/1.7-nm-GaAs superlattice structure under the final Al cell conditions.}
\end{figure} 
\begin{figure}[t]
  \begin{center}
    \psfig{ file=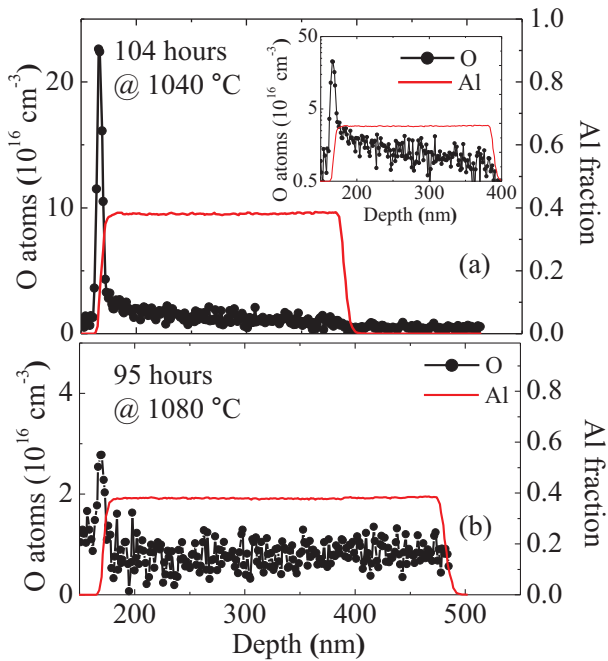, width=0.45\textwidth }
  \end{center}
  \caption{\label{fig3} High-resolution SIMS data for the amount of O in a GaAs/Al$_x$Ga$_{1-x}$As structure for different Al cell bake-out conditions: (a) 104 hour bake at the cell temperature that yields an AlAs deposition rate of 1.33 {\AA}s$^{-1}$,  and (b) that plus an additional 95 hours at 40\textdegree C higher than this temperature. The inset in (a) shows the same data on a logarithmic scale. Note here that the scale is an order of magnitude larger for the case of (a) compared to (b). The intensity for other elements, such as C, Si, Cl, F, Ge, N, P, and S was below the detection limit in both samples.}
\end{figure}
\begin{figure} [t]
  \begin{center}
    \psfig{file=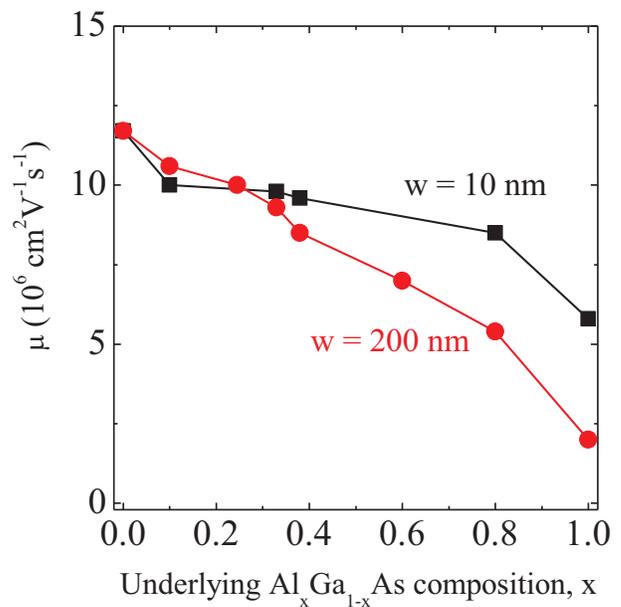, width=0.45\textwidth }
  \end{center}
  \caption{\label{fig4} Measured mobility values of GaAs QWs as a function of the underlying Al$_x$Ga$_{1-x}$As barrier alloy fraction, $x$. The black and red profiles show data for the series of samples with underlying barrier thicknesses of 10 nm and 200 nm, respectively. Here, the bake-out condition for the Al cell used to grow the underlying Al$_x$Ga$_{1-x}$As barrier was 48 hours at 20\textdegree C higher than the temperature that gives a growth rate of 1.33 {\AA}s$^{-1}$.}
\end{figure}

Figure 2 shows the evolution of mobility as a function of underlying Al$_{0.33}$Ga$_{0.67}$As barrier thickness, $w$. As noted earlier and shown in Fig. 1, since these structures are only doped from the top side of the QW, varying conditions beneath the GaAs layer should not significantly alter the carrier density measured in the QW. Indeed, for all the data points shown in Fig. 2, the measured 2DES density was $n\simeq2.1\times10^{11}$ cm$^{-2}$, consistent with the density expected for an Al$_{0.33}$Ga$_{0.67}$As/GaAs heterostructure with $s=80$ nm  when measured after light illumination at $\sim10$ K. It is clear from the data that even though the underlying barrier does not influence the electron density in the GaAs QW, it has a significant impact on the transport mobility. Moreover, this effect is evidently dependent on the thickness of the underlying barrier as well as the purity of the Al source. For example, after the newly-loaded Al cell is outgassed at 1040 \textdegree C for 22 hours, the mobility drops drastically from 9.2$\times10^6$ cm$^{2}$V$^{-1}$s$^{-1}$ when $w=0$ to 1.3$\times10^6$ cm$^{2}$V$^{-1}$s$^{-1}$ when $w=60$ nm. The temperature of 1040 \textdegree C yields an AlAs growth rate of 1.33$\pm 0.02$ {\AA}s$^{-1}$ in our system, which we used to grow Al$_{0.33}$Ga$_{0.67}$As. However, when the Al cell is baked offline for another 104 hours at the same temperature, the mobility only drops to 5.6$\times10^6$ cm$^{2}$V$^{-1}$s$^{-1}$ at $w=60$ nm. As we additionally outgas the Al cell at temperatures  40 \textdegree C and 65 \textdegree C higher than 1040 \textdegree C, the Al source gets cleaner, and the detrimental effect of the underlying barrier becomes apparent only when $w$ is sufficiently thick.

The data in Fig. 2 imply that there must be a build-up of O impurities that can act as scattering sites for electrons as the underlying Al$_x$Ga$_{1-x}$As barrier is being grown. If this were not the case and we assume that the O impurities from the Al source were static, the drop in mobility would be most notable when $w$ is small since this introduces a scattering source nearest to the 2DES in the GaAs QW and then the subsequent drop in mobility would gradually get smaller as $w$ gets larger because it would only generate an additional scattering source further and further away from the 2DES \cite{fnote}. The data measured after bake-out at elevated temperatures show that this hypothesis is simply not true. For these cases, as shown in Fig. 2 the mobility stays at roughly the same value for $10\leq w\leq100$ nm and only shows a noticeable drop at significantly larger $w$.

One model that can account for the behavior in Fig. 2 is the surface segregation of O during the growth of the underlying Al$_{0.33}$Ga$_{0.67}$As barrier and its subsequent discharge into the GaAs QW. In this picture, only a fraction of the O atoms that land on the Al$_{0.33}$Ga$_{0.67}$As barrier are incorporated into the  Al$_{0.33}$Ga$_{0.67}$As material, while another fraction migrates with the moving growth front and accumulates there. When the growth front changes to GaAs, there are no more Al sites for the O atoms to attach to and they are discharged into the GaAs QW. A previous report on the SIMS depth profile of O as $x$ is varied in Al$_x$Ga$_{1-x}$As layers supports this view \cite{SIMS.JCG}. Under these circumstances, a larger barrier thickness is equivalent to having more O impurities at the growth front. This interpretation is reasonably consistent with the data trends observed in Fig. 2, where the drop in mobility is more severe for thicker barriers and the mobility improves as the Al cell is outgassed and fewer O impurities are supplied during growth.

A structure that can test the validity of our hypothesis is one where the underlying Al$_{0.33}$Ga$_{0.67}$As barrier is replaced with a superlattice (SL) consisting of Al$_{0.33}$Ga$_{0.67}$As/GaAs layers. Since each of the GaAs layers in the SL should act as traps for the O impurities accumulated during the growth, we would expect the mobility to improve when the SL is implemented since not as many O impurities should be dumped into the GaAs where the 2DES resides \cite{Sajoto.APL}. We find that indeed this is the case when we replace a 350-nm-Al$_{0.33}$Ga$_{0.67}$As barrier with a 10-nm-Al$_{0.33}$Ga$_{0.67}$As/1.7-nm-GaAs SL with equivalent thickness (350 nm total) after our final bake-out of the Al cell. As shown in Fig. 2, the mobility of the 2DES increases to 11.7$\times10^6$ cm$^{2}$V$^{-1}$s$^{-1}$ from 2.8$\times10^6$ cm$^{2}$V$^{-1}$s$^{-1}$ when the SL is used instead of plain Al$_{0.33}$Ga$_{0.67}$As. This high mobility value is similar to what is observed for cases where $w$ is thin enough that surface segregation is not an issue.

To verify whether surface segregation indeed occurs during the MBE growth of Al$_x$Ga$_{1-x}$As layers, we performed high-resolution SIMS analysis on GaAs/Al$_{0.33}$Ga$_{0.67}$As structures ($x=0.33$) grown after different outgas conditions of the Al cell. Figures 3 (a) and (b) show the SIMS profile for O for two structures grown after the 104-hour bake-out at 1040 \textdegree C and subsequent 95-hour bake-out of the Al cell at 1080 \textdegree C, respectively. We also performed SIMS  for the elements C, Si, Cl, F, Ge, N, P, and S, but their signals were well below the SIMS detection limit \cite{footnote3}. Although undetectable from SIMS, this does not necessarily mean that that these elements are completely absent in the structure, meaning they could still have an effect on the mobility.

In Fig. 3(a), there is an evident build-up of O in the Al$_{0.33}$Ga$_{0.67}$As layer in the growth direction, which culminates in the form of a peak in the SIMS data once in the GaAs layer. Since O is a charged impurity in GaAs \cite{Schneider,Skowronski,Colleoni}, this strongly suggests that our previous hypothesis of surface segregation and discharge of O impurities is correct. A similar trend is observed in Fig. 3(b) with a weaker intensity. In conjunction with the mobility data, this further corroborates our hypothesis as it shows that when the Al cell is outgassed at higher temperatures, the mobility drops less because there are fewer O impurities segregated toward the GaAs 2DES. The marked difference observed in the signal intensity between the two samples in Fig. 3 also rules out the possibility of the data being an artifact of a matrix effect during the SIMS measurement \cite{Matrix}, since the beam etches through an identical GaAs/Al$_{0.33}$Ga$_{0.67}$As structure for both cases.

We would like to add that the weaker signal in Fig. 3(b) compared to Fig. 3(a) is also consistent with the fact that the H$_2$O peak in our residual gas analyzer decreased from $\simeq5\times10^{-12}$ Torr when it was first used in growth to $\simeq1\times10^{-12}$ Torr when the Al cell was outgassed for 95 hours at 1080 \textdegree C. This was further reduced to $\simeq5\times10^{-13}$ Torr when the Al cell was outgassed for 48 hours at 1105 \textdegree C.

Given these observations, it is also interesting to evaluate the influence of the barrier alloy fraction $x$ on the surface segregation of O impurities during growth. Figure 4 shows the measured mobility values for samples where $x$ was varied for the underlying Al$_x$Ga$_{1-x}$As for the two cases of $w=10$ nm and 200 nm. Here, we used {\it a different} Al cell that was outgassed at 1050 \textdegree C for 48 hours after it was newly loaded into the MBE chamber. In this cell, the temperature that yields an AlAs growth rate of 1.33 {\AA}s$^{-1}$ was 1030\textdegree C. We fixed the temperature of this Al cell during growth to be 1030 \textdegree C for all the samples to ensure that O impurity outgassing from the Al cell was constant. The top Al$_{0.33}$Ga$_{0.67}$As barrier was grown using a Ga cell with a deposition rate of 2.83 {\AA}s$^{-1}$, while the Al alloy fraction $x$ of the underlying barrier was tuned by varying the temperature (and hence deposition rate) of another Ga cell \cite{foot2}. As discussed earlier, the $w=10$ nm barrier series was grown to distinguish the effect of introducing a barrier and defining a QW from the effect surface segregation of O impurities. In Fig. 4, although there is a weak dependence on $x$ when $w=10$ nm, comparing it to the case when $w=200$ nm, it is clear that the surface segregation of O impurities becomes more pronounced as $x$ increases. This suggests that the amount of surface-segregated O is directly proportional to the Al concentration in the vicinity of the growth front given a constant flux of O atoms (fixed Al cell temperature in our case). 

In conclusion, we have shown that the purity of the Al source has a significant impact on the electron mobility of a high-quality, modulation-doped GaAs 2DES. We quantify this effect by comparing electron mobilities of a series of GaAs 2DESs where either the thickness $w$ or barrier alloy fraction $x$ of the underlying barrier is varied as we outgas a newly-loaded Al cell in our MBE. Even after substantial purification of the Al cell, a noticeable drop is observed in the mobility of the 2DES when $w$ or $x$ is large, which we attribute to the surface segregation and discharge of impurities from the underlying Al$_x$Ga$_{1-x}$As barrier into the GaAs. High-resolution SIMS results confirm this model, and reveal that the primary culprit is O. 

Our results imply that, in addition to having a clean Ga source and extremely good vacuum, a clean Al source is essential to obtain high quality 2DESs in GaAs QWs, especially in structures including a barrier that is thick or has high $x$. Examples of samples with such features are low-density QW structures which require thick Al$_x$Ga$_{1-x}$As spacers, or double-QW structures with negligible interlayer tunneling and strong interlayer Coulomb interaction; the latter comprise a high-$x$ barrier between the two QWs. Another, extreme example is a structure where the 2DES resides in a pure AlAs QW \cite{Lay.PRB.1993,DePoortere.APL.2002,Shayegan.PSS.2006,Chung.PRM.2017}. Indeed, we have been able to fabricate AlAs 2DESs with unprecedented quality and high mobility ($>2\times10^{6}$ cm${^2}$V$^{-1}$s${^-1}$) when we use a sufficiently outgassed Al cell to grow the structure; details will be described in a future publication. It is also noteworthy that the scheme reported here could be used to quantitatively assess the purity of an Al cell in an MBE chamber by increasing the barrier thickness or the Al alloy fraction until a drop in mobility is observed.

\begin{acknowledgments}
We acknowledge support through the NSF (Grants DMR 1709076 and ECCS 1508925) for measurements, and the NSF (Grant MRSEC DMR 1420541) and the Gordon and Betty Moore Foundation (Grant GBMF4420) for sample fabrication and characterization.
 \end{acknowledgments}


\begin{thebibliography}{99}
\bibitem{fnote10} The Ga source was purified in a prior series of outgassing tests. The high purity of the Ga source is shown by the fact that the mobility of the 2DES hosted in a GaAs heterostructure was already $9.2\times10^6$ cm$^2$V$^{-1}$s$^{-1}$ even when the Al cell was only baked out briefly at 1040\textdegree C (See Fig. 2).
\bibitem{fnote} It is important to note here that the initial drop in mobility when comparing the cases of $w=0$ and $w=10$ nm at higher outgassing temperatures is most likely because the 2DES in the GaAs layer transitions from a heterostructure to a true QW, rather than due to impurities being introduced in proximity to the GaAs. This is consistent with the mobility staying at roughly the same value for $10\leq w\leq100$ nm at these temperatures.
\bibitem{footnote3} The following are the detection limits for each element in GaAs for the SIMS analysis performed in this work, in units of atoms/cm$^3$: C $\simeq 6\times10^{14}$, O $\simeq 3\times10^{15}$, Si $\simeq 4\times10^{14}$, F $\simeq 3\times10^{13}$, H $\simeq 1\times10^{17}$, Ge $\simeq 5\times10^{15}$, Cl $\simeq 7\times10^{13}$, N $\simeq 2\times10^{17}$, P $\simeq 4\times10^{16}$, and S $\simeq 2\times10^{14}$. In Al$_{0.33}$Ga$_{0.67}$As, the detection limits for O, C, and Si are higher: $\simeq 1\times10^{16}$, $\simeq 5\times10^{15}$, and $\simeq 1\times10^{15}$, respectively.
\bibitem{foot2} In the case of $x=0.1$, fixing the Al cell temperature at 1030 \textdegree C  yielded a total  Al$_x$Ga$_{1-x}$As growth rate higher than 5 {\AA}s$^{-1}$ which could be detrimental to sample quality. Therefore for the $x=0.1$ sample we used a lower Al cell temperature of 960 \textdegree C (AlAs deposition rate 0.33 {\AA}s$^{-1}$).
\end{thebibliography}
\end{document}